\documentclass[twocolumn,nofootinbib,amsmath,showkeys]{revtex4}
\usepackage{graphicx}

\begin{document}

\title{Scale Invariance without Inflation?}

\author{C. Armend\'ariz-Pic\'on}
\affiliation{Enrico  Fermi Institute and  Department of  Astronomy and
  Astrophysics, \\University of Chicago.}

\author{Eugene  A.  Lim}
\affiliation{Center for
  Cosmological Physics and  Department of Astronomy  and Astrophysics, \\
  University of Chicago.}

\begin{abstract}
  We propose  a new  alternative mechanism to  seed a  scale invariant
  spectrum of  primordial density perturbations that does  not rely on
  inflation.   In our scenario,  a perfect  fluid dominates  the early
  stages of  an expanding, non-inflating universe.   Because the speed
  of sound of  the fluid decays, perturbations are  left frozen behind
  the sound horizon,  with a spectral index that  depends on the fluid
  equation of state.   We explore here a toy  model that realizes this
  idea.  Although the model  can explain an adiabatic, Gaussian, scale
  invariant spectrum of primordial perturbations, it turns out that in
  its simplest  form it cannot  account for the observed  amplitude of
  the primordial density perturbations.
\end{abstract}
\keywords{cot, inf}

\maketitle

\section{Introduction}
During the last decade, several  experiments have probed the nature of
the primeval perturbations that gave rise to the anisotropies observed
in the  universe \cite{WMAP}.  The  results of these  measurements are
consistent  with an  adiabatic,  Gaussian and  nearly scale  invariant
primordial  spectrum of  perturbations, as  predicted by  the simplest
models of inflation \cite{MukhanovChibisov}.   Due to the central role
that the origin  of these perturbations plays in  our understanding of
the  early universe,  it  is  important to  ascertain  to what  extent
alternatives  to seed  such a  spectrum exist.   In the  context  of a
universe  dominated by  a  single canonical  scalar  field coupled  to
general   relativity,   this    issue   has   been   investigated   in
\cite{GrKhStTu}.   In a contracting  universe, there  are two  ways to
seed  a  scale  invariant  spectrum  \cite{alternatives},  and  in  an
expanding universes only  a stage of de Sitter  inflation can generate
it.  Other  alternatives that  relax some of  the assumptions  made in
\cite{GrKhStTu}         have         been         discussed         in
\cite{BrandenbergerHo,HollandsWald,  AvelinoMartins,  Oaknin}.  It  is
fair to say though, that the only presently known, widely accepted way
to seed  a scale invariant  spectrum of perturbations in  an expanding
universe requires a stage of de Sitter inflation.

During inflation,  the physical length of a  perturbation grows faster
than the  Hubble radius $H^{-1}$.   Hence, because for a  scalar field
the sound of speed is one, initially sub-horizon sized modes cross the
\emph{sound} horizon and subsequently  freeze.  Note that the crossing
of the sound horizon $c_s H^{-1}$, rather than the Hubble radius,
is  what leads  to  the causal  seeding  of a  primordial spectrum  of
density perturbations.

In  this paper  we  propose  an alternative  seeding  mechanism in  an
\emph{expanding}  universe that  does not  require inflation.   In our
scenario,  modes  cross the  sound  horizon  because  the sound  speed
decreases  sufficiently  fast.   The  amplitude of  the  perturbations
depends on the  values of the Hubble parameter and  the sound speed at
the time of crossing.  If  the sound speed decays appropriately, it is
also possible  to seed a scale  invariant spectrum.  In  order to push
the seeded perturbations to  super-Hubble scales, it is necessary that
a period of  inflation follows the seeding. However,  let us emphasize
that   the  seeding   and  inflation   are  two   distinct,  unrelated
cosmological phases; the nature  of the seeded spectrum is independent
of the properties of the inflationary stage, which does not have to be
de Sitter-like.

We discuss  a particular realization  of the scenario outlined  above. 
Although the particular model we discuss can successfully seed a scale
invariant  spectrum of  density perturbations,  it turns  out  that it
cannot  account  for the  observed  amplitude.   Specifically, if  one
requires the amplitude  of the seeded perturbations to  agree with the
one observed  in CMB anisotropies, our  model can seed  only about two
decades in $k$ space (instead of  the required three to four).  On the
other hand, if  one requires the seeding to  span about three decades,
then the amplitude of the  seeded perturbations has to be smaller than
the observed one.  The only way to escape this conclusion is to assume
that  between seeding  and  observation the  amplitude  of the  seeded
spectrum is boosted to the observed value by a different mechanism.

\section{Formalism}
Consider  a spatially  flat FRW  universe dominated  by  an isentropic
perfect fluid.  We would like  to study scalar perturbations in such a
universe \cite{MuFeBr}, so we shall deal with the perturbed metric (in
longitudinal gauge)
\begin{equation}
ds^2=a^2(\eta)\left[(1+2\Phi)d\eta^2-(1-2\Phi)d\vec{x}^2\right].
\end{equation}
The  evolution of  the scale  factor is  determined by  the background
Einstein equations
\begin{equation}\label{eq:Einstein}
\mathcal{H}^2=a^2\rho\quad \text{and}\quad
\mathcal{H}'=-\frac{\mathcal{H}^2}{2}(1+3w),
\end{equation}
where $\mathcal{H}=a'/a$  and prime denotes a  derivative with respect
to conformal time $\eta$.  We work in units where Newton's constant is
${G=3/(8\pi)}$ and,  as usual, we  have defined the equation  of state
parameter $w=p/\rho$. The pressure of  the fluid is $p$, and $\rho$ is
its energy density. The Hubble parameter is $H=\mathcal{H}/a$.

The evolution of the scalar perturbations can be described in terms of
the  Mukhanov variable  $v$,  which  is a  linear  combination of  the
gravitational   potential   $\Phi$   and   the   fluid   perturbations
\cite{MuFeBr,GarrigaMukhanov}.  The Fourier components of the Mukhanov
variable $v_k$ obey the simple differential equation
\begin{equation}\label{eq:motion}
  v_k''+\left(c_s^2 k^2-\frac{z''}{z}\right)v_k=0,
\end{equation} 
where the variable $z$ is given by
\begin{equation}\label{eq:z}
  z=\frac{a (1+w)^{1/2}}{c_s},
\end{equation}
and the  speed of sound of the isentropic fluid is
\begin{equation}\label{eq:cs}
  c_s^2\equiv \frac{dp}{d\rho}=w-\frac{w'}{3\mathcal{H}(1+w)}.
\end{equation}
Note that for  isentropic fluids, the equation of  state and the speed
of   sound   are   linked    by   the   fluid   equation   of   motion
$\rho'+3\mathcal{H}\rho(1+w)=0$.

Whereas the variable $v$ proves to be useful to study the quantization
and the evolution of the perturbations, the Bardeen variable
\begin{equation}\label{eq:Bardeen}
  \zeta=\frac{v}{z}=
  \frac{2}{3}\frac{\mathcal{H}^{-1}\Phi'+\Phi}{1+w}+\Phi
\end{equation}
turns to be more convenient  to compute observable quantities, such as
the temperature anisotropies in the CMB radiation.  These anisotropies
can be characterized by the the power spectrum
\begin{equation}\label{eq:powerdef}
  \mathcal{P}_\zeta=\frac{k^3}{4\pi^2}|\,\zeta_k|^2,
\end{equation}
which is  a measure  of the mean  square fluctuations of  the variable
$\zeta$ on comoving scales $1/k$.  The variable $\zeta$ is also useful
because it can be directly  related to the perturbations in the energy
density. In fact,  in cases where $\Phi$ is  constant, it follows from
Eq.   (\ref{eq:Bardeen}) that ${\zeta_k\sim\Phi_k}$.   Using Poisson's
equation (for modes smaller than the Hubble radius) we then find
\begin{equation}\label{eq:deltarho}
  \mathcal{P}_{\delta\rho/\rho}\sim 
  \frac{k^4\,\mathcal{P}_\zeta}{a^4\,\rho^2},
\end{equation}
which connects  the inhomogeneities in $\zeta$  to the inhomogeneities
in the fractional density contrast.

Let  us  turn  our  attention   to  the  equation  of  motion  of  the
perturbations,   Eq.   (\ref{eq:motion}).    By  definition,   in  the
short-wavelength regime  ${c_s^2 k^2\gg z''/z}$. If $c_s$  and $w$ are
constant, the  last condition  translates into $c_s^2  k^2\gg a^2H^2$,
which means  that modes are smaller  than the sound  horizon.  In this
regime, an approximate solution of Eq.  (\ref{eq:motion}) is
\begin{equation}\label{eq:adiabatic}
  v_k(\eta)=\frac{1}{\sqrt{c_s(\eta) k}}
  \exp \left(-i\int^\eta c_s(\tilde{\eta})\, k\, d\tilde{\eta}\right).
\end{equation}
This WKB-like solution is valid as long as the change in the frequency
$c_s k$ is adiabatic,
\begin{equation}\label{eq:adiabaticconditions}
  c_s^2 k^2\gg \frac{c_s''}{c_s}\quad \text{and} \quad
  c_s^2 k^2\gg \left(\frac{c_s'}{c_s}\right)^2.
\end{equation}
Then,  the approximate  solution  (\ref{eq:adiabatic}) corresponds  to
zeroth-order adiabatic vacuum initial conditions \cite{BirrellDavies}.
In  the  long-wavelength regime,  $c_s^2  k^2\ll  z''/z$,  one of  the
solutions of Eq.  (\ref{eq:motion}) is proportional to $z$.  Hence, if
$v\propto z$  is the dominant solution of  Eq.  (\ref{eq:motion}), for
long-wavelength modes the variable $\zeta$ is conserved.  This remains
true as long as there are no entropy perturbations.

\section{Decaying sound speed}
In order to causally seed  a spectrum of perturbations it is necessary
that modes  that initially lie  in the short-wavelength  regime, where
natural initial conditions exist, enter the long-wavelength regime and
subsequently freeze. This occurs only if
\begin{equation}\label{eq:seeding}
  \left(\frac{z''}{c_s^2 z}\right)'>0.
\end{equation}

One  way to  satisfy condition  (\ref{eq:seeding}) is  to  assume that
$c_s=1$ and $w<-1/3$.  Then, $z''/z\sim a^2 H^2$ increases while $c_s$
remains constant.   In physical terms, this means  the physical length
of a  mode, $a/k$, grows faster  than the Hubble  radius $H^{-1}$, the
condition that  singles out inflation.  Recall that  for a (canonical)
scalar field $c_s$ is indeed equal one \cite{GarrigaMukhanov}.

There  are  nevertheless  additional   ways  to  satisfy  the  seeding
condition (\ref{eq:seeding}).   The main idea  of our paper is  that a
rapidly changing  speed of sound can  also result in the  seeding of a
scale invariant spectrum.  Even  if ${z''/z\sim a^2 H^2}$ decreases in
time, if $c_s$ changes fast enough it is possible to satisfy condition
(\ref{eq:seeding}). Physically,  this means that the length  of a mode
grows  faster than  the  \emph{sound} horizon  $c_s  H^{-1}$.  In  the
following we present a model that realizes this idea.  For simplicity,
we  assume  here  that  the  fluid  that  dominates  the  universe  is
isentropic.   In that  case,  the  sound speed  is  determined by  the
equation of state, which in turn is determined by the expansion of the
universe. Hence, the  free parameters of our simple  model are tightly
constrained.   By considering  non-isentropic fluids  or non-canonical
scalar fields, one could avoid these restrictions.

\section{A simple model}

Suppose that  the early universe  is dominated by an  isentropic fluid
with polytropic equation of state
\begin{equation}\label{eq:pofrho}
  p=w_* \rho_* \left(\frac{\rho}{\rho_*}\right)^{1+\alpha/3}.
\end{equation}
Here $\alpha$  is a free  parameter that characterizes  the polytropic
index and  $w_*$ is the value  of the equation of  state parameter for
$\rho=\rho_*$.  Perfect fluids with polytropic equations of state have
been long and widely considered in astrophysics \cite{polytropic}.

If $w_*\ll1$, for energy densities  below $\rho_*$ the pressure $p$ is
negligible.   Under the assumption  $p\ll \rho$,  integration of  Eqs. 
(\ref{eq:Einstein}) yields
\begin{equation}\label{eq:aandH}
  a= a_* \left(\frac{\eta}{\eta_*}\right)^2,\quad 
\mathcal{H}=\frac{2}{\eta},
\end{equation}
which is of course how  a dust-dominated universe evolves in conformal
time.  Although  the equation of  state parameter and the  sound speed
are small, they are  not exactly zero.  Using Eqs.  (\ref{eq:pofrho}),
(\ref{eq:Einstein}) and (\ref{eq:cs}) we find
\begin{equation}\label{eq:cseta}
  w=w_* \left(\frac{\eta}{\eta_*}\right)^{-2\alpha},\quad
  c_s^2=\left(1+\frac{\alpha}{3}\right) w(\eta).
\end{equation}
Finally,  substituting  these expressions  into  Eq.  (\ref{eq:z})  we
arrive at
\begin{equation}\label{eq:z''z}
  \frac{z''}{z}=\frac{2+3\alpha+\alpha^2}{4}\mathcal{H}^2.
\end{equation}

Therefore, the seeding  condition Eq.  (\ref{eq:seeding}) is satisfied
only  if $c_s$  decays fast  enough, $\alpha>1$.   The origin  of this
condition  can be also  interpreted in  physical terms.   The physical
length of a  mode is $\lambda=a/k$, and the size  of the sound horizon
is  $c_s H^{-1}$.  Requiring  the physical  length of  a mode  to grow
faster than  the sound horizon  also yields $\alpha>1$.  Note  that if
$\alpha>3$, both speed of sound and sound horizon decay.

With $z''/z$  given by Eq.  (\ref{eq:z''z})  and $c_s^2$ given  by Eq. 
(\ref{eq:cseta}) the  differential equation (\ref{eq:motion}) has
the solution
\begin{equation}\label{eq:solution}
  v_k=\eta^{1/2}\left[A_k H_\nu\left(\frac{c_s\, k}{\alpha-1}\eta\right)
    +B_k H^*_\nu\left(\frac{c_s\, k}{\alpha-1}\eta\right)\right],
\end{equation}
where $H_\nu$ is the Hankel function of the first kind and
\begin{equation}\label{eq:nu}
  \nu=\frac{2\alpha+3}{2(\alpha-1)}.
\end{equation}
The coefficients $A_k$ and  $B_k$ are integration constants determined
by the initial conditions.  For modes well within the short-wavelength
regime, the adiabaticity conditions (\ref{eq:adiabaticconditions}) are
satisfied. Setting
\begin{equation}
  A_k=\sqrt{\frac{\pi}{2(\alpha-1)}}, \quad B_k=0
\end{equation}
the  solution  (\ref{eq:solution})  approaches  the  adiabatic  vacuum
(\ref{eq:adiabatic}).  Substituting Eq.   (\ref{eq:solution}) into Eq. 
(\ref{eq:powerdef})  we find  that in  the long-wavelength  regime the
power-spectrum $\mathcal{P}_\zeta$ is time-independent.  Expressing it
in terms of the values of $H$  and $c_s$ at the time a reference scale
$k_*$ crosses the sound horizon, $c_s^* k_*=a_* H_*$, results into
\begin{equation}\label{eq:power}
  \mathcal{P}_\zeta=\frac{\Gamma^2(\nu)\cdot (\alpha-1)^{2\nu-1}}{4\pi^3}
  \cdot\frac{H_*^2}{c_s^*}\left(\frac{k}{k_*}\right)^{n_s-1},
\end{equation}
where $\nu$ is determined by Eq.  (\ref{eq:nu}) and the spectral index
is given by
\begin{equation}
  n_s-1=\frac{\alpha-6}{\alpha-1}.
\end{equation}
Hence, for $\alpha>6$ the spectrum is blue, for $\alpha=6$ it is scale
invariant\footnote{Although   this   statement    is   true   in   our
  approximation $w\to0$,  the arbitrarily small  deviations from $w=0$
  will  cause deviations from  scale invariance  even if  $\alpha=6$.} 
and  for $\alpha<6$  it  is red.   In  particular, one  can imagine  a
situation where $\alpha$ is  mildly $\rho$-dependent and increasing as
a function  of $\rho$.  In such  a case, the spectral  index would run
from  $n_s>1$  to $n_s<1$  at  smaller  scales,  as hinted  by  recent
observations \cite{WMAP}.  Because  the seeded perturbations originate
from  quantum fluctuations  of  the  fluid, they  are  expected to  be
Gaussian.

\subsection{Observational  constraints}

Large  scale  structure  and  CMB measurements  probe  the  primordial
spectrum on scales  that range from the size  of today's Hubble radius
to  scales about  $10^3$ times  smaller. Presently,  these  probes are
consistent  with  an   adiabatic,  Gaussian,  nearly  scale  invariant
spectrum with amplitude  $\mathcal{P}_\zeta\sim 10^{-10}$ \cite{WMAP}. 
In  this  section   we  explore  whether  our  model   is  capable  of
successfully  accounting  for these  primordial  anisotropies and  how
these observations constrain and affect our model.

Suppose that the seeding of  perturbations we have described lasts for
$\Delta  N_s$  $e$-folds.   The  minimal  value  of  $\Delta  N_s$  is
determined by  the requirement that our  mechanism seeds perturbations
spanning  three decades in  $k$ space. Because  modes are  seeded when
they cross the sound horizon, $c_s k=a H$, using Eqs. (\ref{eq:aandH})
and (\ref{eq:cseta})  one thus finds  that the number of  $e$-folds of
seeding has to satisfy
\begin{equation}\label{eq:e-folds}
\Delta N_s\geq \frac{D \log 10}{1+\alpha/2-3/2},
\end{equation}
where  $D$ is  the number  of cosmologically  relevant decades  in $k$
space (about three).   After the seeding, we assume  that the universe
evolves with  constant equation of state $\widetilde{w}$  for a period
of  $\Delta  N_{\widetilde{w}}$  $e$-folds.   Next,  the  universe  is
reheated  at  $a_{rh}$  and  it  subsequently  undergoes  a  radiation
dominated phase until equipartition $a_{eq}$.  Following equipartition
the  universe   becomes  matter  dominated  until   today  $a_0$  (for
simplicity we neglect the recent stage of cosmic acceleration).  Thus,
at the beginning of the seeding, the physical size of a comoving scale
$k_*$ was
\begin{equation}
  \lambda_*=\frac{a_0}{k_*}\frac{a_{eq}}{a_0}\frac{a_{rh}}{a_{eq}}
  \exp(-\Delta N_{\widetilde{w}}) \exp(-\Delta N_s).
\end{equation}
On the other hand, the value of the Hubble parameter at that time was 
\begin{equation}\label{eq:Hstar}
  H_*=H_0\left(\frac{a_0}{a_{eq}}\right)^{3/2}
  \left(\frac{a_{eq}}{a_{rh}}\right)^2 
  e^{3\Delta N_{\widetilde{w}} (1+\widetilde{w})/2} e^{3\Delta N_s/2}.
\end{equation}
Using the last  two equations, one can compute the  value of the speed
of sound at  the time the present Hubble  radius $k_*=a_0 H_0$ crossed
the sound horizon during seeding, $c_s^* H_*^{-1}=\lambda_*$,
\begin{equation}\label{eq:csstar}
  c_s^*=\left(\frac{a_0}{a_{eq}}\right)^{1/2}
  \left(\frac{a_{eq}}{a_{rh}}\right)
  e^{(1+3\widetilde{w})\Delta N_{\widetilde{w}}/2} e^{\Delta N_s/2}.
\end{equation}
Equipartition occurred at $a_0/a_{eq}\approx 3\cdot 10^3$, and because
reheating  has to  happen  before nucleosynthesis,  $a_{eq}/a_{rh}\geq
10^8$.  Since the  speed of sound and the  equation of state parameter
are   related   by   Eq.    (\ref{eq:cseta}),  inspection   of   Eq.   
(\ref{eq:csstar}) shows  that in order  to prevent a violation  of the
condition $w\ll1$, we have to  assume that the seeding was followed by
a stage of inflation, $\widetilde{w}<-1/3$.

\begin{figure}
  \begin{center}
    \includegraphics{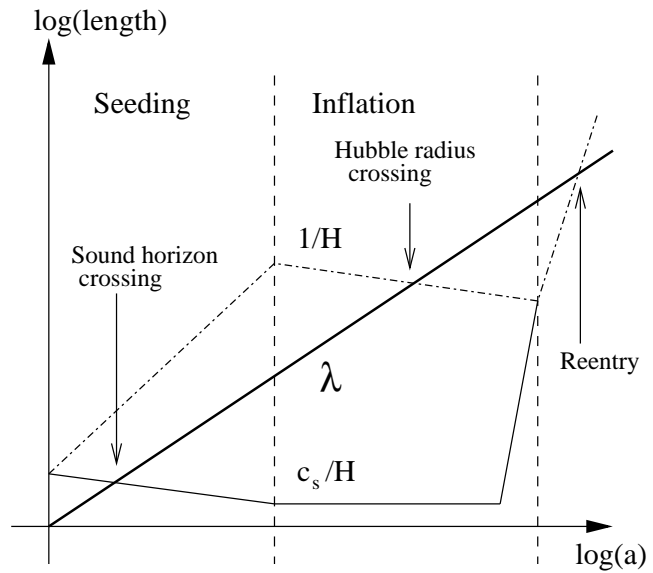}
    \caption{A schematic plot of the evolution of the different
      length  scales in  our scenario.   During seeding,  the physical
      length $\lambda$ (thick line)  crosses the sound horizon $c_s/H$
      (continuous line).  Later, during inflation, $\lambda$ exits the
      Hubble radius $1/H$ (dot-dashed line).
      \label{fig:scales}}
  \end{center}
\end{figure}

The evolution of  a particular comoving scale $k$  in time is depicted
in Fig.  \ref{fig:scales}.  Observe that in our model,  the seeding of
the scalar perturbations and the inflationary epoch are decoupled from
each other.   The only purpose of  the inflationary stage  is to allow
for a small speed of sound  during the seeding and to push the already
seeded  spectrum of  perturbations to  scales larger  than  the Hubble
radius.   Gravitational waves  are  seeded only  during  the stage  of
inflation.  Because  the amplitude of the tensor  perturbations at the
time a given  scale crosses the Hubble radius  is proportional to $H$,
the tensor  to scalar ratio  at $k_*=a_0 H_0$  is, up to  constants of
order one,
\begin{equation}
  \frac{\mathcal{P}_h}{\mathcal{P}_\zeta}\Bigg|_{k_*}\sim
  (c_s^*)^{-\frac{5+3\widetilde{w}}{1+3\widetilde{w}}}\cdot 
  \exp\left(-\frac{6\widetilde{w}}{1+3\widetilde{w}}\Delta N_s\right). 
\end{equation} 
Since $c_s^*$ is small, tensor perturbations are suppressed.  The same
argument  applies   to  scalar  perturbations   generated  during  any
inflationary regime  driven by a component different  from the seeding
fluid.

Inflation  might be  driven  by  the same  component  that causes  the
seeding, e.g.  by a k-field  \cite{k-inflation}, or it might be driven
by  a second  component, like  a conventional  scalar.  In  the former
case, in order  to preserve the seeded spectrum,  it is important that
perturbations do not reenter the sound horizon during inflation.  This
is possible  during k-inflation,  because the speed  of sound  and the
equation of state parameters are two arbitrary independent quantities.
In  the latter case,  it is  important that  no entropic  component be
generated  during inflation.   Just as  for gravitational  waves, such
entropic  perturbations are expected  to be  negligible. Nevertheless,
the   presence  of   a   second  component   could  have   non-trivial
consequences,  such  as  for  instance,  a  boost  in  the  primordial
amplitude  of the  perturbations  due to  parametric resonance  during
reheating \cite{FinelliBrandenberger}.

It  turns  out  that  the  validity of  perturbation  theory  severely
constrains our  model. Using Eq.  (\ref{eq:deltarho})  one can compute
the value of  $\delta\rho/\rho$ at the time of  sound horizon crossing
for  the  last seeded  mode\footnote{The  constraint  that follows  by
  considering this  mode is stronger  than the one requiring  that the
  density  perturbation  for  the  mode  $k_*$ remain  in  the  linear
  regime.},
\begin{equation}
  \mathcal{P}_{\delta\rho/\rho}=
  \frac{\mathcal{P}_\zeta}{{c_s^*}^4 \exp\left(-2\alpha\Delta N_s\right)}.
\end{equation}
The      validity       of      perturbation      theory      requires
$\mathcal{P}_{\delta\rho/\rho}<1$,   and   self-consistency   of   our
derivation $c_s^*<1$. Therefore we find
\begin{equation}
  \Delta N_s<\frac{\log\mathcal{P}_\zeta^{-1}}{2\alpha}.
\end{equation}
The  amplitude   of  the  primordial  fluctuations   observed  in  CMB
temperature  anisotropies   implies  $\mathcal{P}_\zeta\sim  10^{-10}$
\cite{WMAP}.   Setting $\alpha=6$  ($n_s=1$) in  the last  equation we
thus  find  $\Delta  N_s<2$,   which  is  somewhat  smaller  than  the
observationally required  $\Delta N_s\approx 3$  that follow from  Eq. 
(\ref{eq:e-folds}) (for $D=3$).

\subsection{A Microscopic Description}
Although Eq.   (\ref{eq:pofrho}) uniquely specifies  the perfect fluid
model,  it might be  desirable to  cast our  model into  a microscopic
formulation.  Consider a scalar field $\varphi$ with Lagrangian
\begin{equation}
  L=p(X), \quad \text{where}\quad
  X=\frac{1}{2}g^{\mu\nu}
  \frac{\partial\varphi}{\partial x^\mu}
  \frac{\partial\varphi}{\partial x^\nu}.
\end{equation}
Here, $p$  is an  as of  yet unspecified, arbitrary  function of  $X$. 
Such  a  Lagrangian   describes  a  k-field  \cite{k-inflation}.   For
timelike field  gradients, the energy  momentum tensor of  the k-field
has  perfect fluid  form, with  pressure  given by  $p(X)$ and  energy
density given by ${\rho(X)=2X  dp/dX-p}$.  Because the Lagrangian only
depends  on $X$, in  this particular  case the  k-field behaves  as an
isentropic fluid.  Using the previous expression for $\rho(X)$ one can
derive  a differential  equation  for  the $p(X)$  that  leads to  the
desired equation of state (\ref{eq:pofrho}).  The solution is
\begin{equation}
  p(X)=w_* \rho_* \left[1+\frac{\alpha}{2w_*(3+\alpha)}
      \log\left(\frac{X}{X_*}\right)\right]^{1+3/\alpha},
\end{equation}
where $p\ll \rho$ has been assumed.

\section{Conclusions}

We have  presented a  novel alternative mechanism  to causally  seed a
spectrum of  density perturbations in  an expanding universe.   In our
scenario,  the decaying  sound speed  of an  isentropic  perfect fluid
causes perturbations to be left  frozen behind the sound horizon, with
a spectrum  that depends  on the  decay rate of  the sound  speed.  In
order for the seeded spectrum to correspond to cosmologically relevant
scales, an inflationary stage has  to follow the phase of perturbation
seeding.  Therefore, the seeding and the spectrum of the perturbations
and the inflationary stage are distinct unrelated cosmological phases.
Since  the   inflationary  stage   happens  after  the   seeding,  the
gravitational  waves  generated  during  the  inflationary  stage  are
suppressed compared to scalar perturbations.

Although  the  particular model  we  have  discussed can  successfully
explain   a   scale   invariant   spectrum  of   Gaussian,   adiabatic
perturbations on observable scales, it cannot account for its observed
amplitude in the cosmologically relevant window of about three decades
in $k$ space.   The only way we are aware of  to avoid this conclusion
is  to assume that  parametric resonance  during reheating  boosts the
amplitude   of  the   primordial  spectrum   to  its   observed  value
\cite{FinelliBrandenberger}.  At  this point  one should also  bear in
mind  that we  have explored  only one  particular realization  of our
scenario.  In  fact, we are not  aware of any  physical principle that
would  prevent other  realizations  to also  account  for the  correct
amplitude in a larger cosmological window.

Why  should  we  consider  a  stage  of  pre-inflationary  seeding  if
perturbations can be generated  during inflation anyway?  It turns out
that in some  cases it might be useful to decouple  the seeding of the
perturbations  from  the  stage  of inflation  \cite{curvaton}.   This
decoupling acquires  particular relevance in  the light of  the recent
signs of a running spectral  index \cite{WMAP}.  Such a running can be
accommodated by  conventional inflationary models only at  the cost of
spoiling their simplicity and economy \cite{ChShTr}.  If in the era of
precision   cosmology  one   is  forced   to  replace   simplicity  by
phenomenological success, less simple alternatives to the conventional
models might be attractive too.

On  the  other  hand, our  results  can  be  also interpreted  from  a
different  perspective:  They display  how  many  hurdles  one has  to
overcome in  order to seed a phenomenologically  realistic spectrum of
perturbations without inflation.

\begin{acknowledgments}
  We thank R.  Brandenberger, Sean  Carroll, Wayne Hu and V.  Mukhanov
  for useful  comments and discussions.   CAP was supported by  the US
  DOE grant  DE-FG02-90ER40560, and  EAL was supported  by the  US DOE
  grant DE-FG02-90ER40560 and the Packard foundation.
\end{acknowledgments}


\begin{thebibliography}{99} 
  
\bibitem{WMAP}  D.~N.~Spergel  {\it et  al.},  ``First Year  Wilkinson
  Microwave  Anisotropy Probe  (WMAP)  Observations: Determination  of
  Cosmological Parameters,'' arXiv:astro-ph/0302209.
  
\bibitem{MukhanovChibisov}V.~F.~Mukhanov and G.~V.~Chibisov, ``Quantum
  Fluctuation And  Nonsingular Universe,''  JETP Lett.\ {\bf  33}, 532
  (1981).

\bibitem{GrKhStTu}   S.~Gratton,   J.~Khoury,   P.~J.~Steinhardt   and
  N.~Turok,   ``Conditions  for  generating   scale-invariant  density
  perturbations,''      arXiv:astro-ph/0301395. 
  
\bibitem{alternatives}  J.~Khoury,  B.~A.~Ovrut, P.~J.~Steinhardt  and
  N.~Turok,  ``Density  perturbations  in  the  ekpyrotic  scenario,''
  Phys.\  Rev.\ D  {\bf 66},  046005 (2002).   [arXiv:hep-th/0109050]. 
  F.~Finelli   and  R.~Brandenberger,   ``On  the   generation   of  a
  scale-invariant spectrum  of adiabatic fluctuations  in cosmological
  models with a  contracting phase,'' Phys.\ Rev.\ D  {\bf 65}, 103522
  (2002) [arXiv:hep-th/0112249].
 
\bibitem{BrandenbergerHo}      R.~Brandenberger      and     P.~M.~Ho,
  ``Noncommutative spacetime, stringy spacetime uncertainty principle,
  and density  fluctuations,'' Phys.\ Rev.\ D {\bf  66}, 023517 (2002)
  [AAPPS Bull.\ {\bf 12N1}, 10 (2002)] [arXiv:hep-th/0203119].
  
\bibitem{HollandsWald} S.~Hollands and R.~M.~Wald, ``An alternative to
  inflation,''   Gen.\   Rel.\    Grav.\   {\bf   34},   2043   (2002)
  [arXiv:gr-qc/0205058].   L.~Kofman,    A.~Linde   and   V.~Mukhanov,
  ``Inflationary theory and  alternative cosmology,'' JHEP {\bf 0210},
  057 (2002) [arXiv:hep-th/0206088].
  
\bibitem{AvelinoMartins}P.~P.~Avelino  and C.~J.~Martins, ``Primordial
  adiabatic  fluctuations from cosmic  defects,'' Phys.\  Rev.\ Lett.\ 
  {\bf 85}, 1370 (2000) [arXiv:astro-ph/0002413].
  
\bibitem{Oaknin}   D.~H.~Oaknin,  ``Spatial  correlations   of  energy
  density  fluctuations in  the fundamental  state of  the universe,''
  arXiv:hep-th/0305068.

\bibitem{MuFeBr}        V.~F.~Mukhanov,        H.~A.~Feldman       and
  R.~H.~Brandenberger,   ``Theory  Of   Cosmological  Perturbations,''
  Phys.\ Rept.\ {\bf 215}, 203 (1992).
  
\bibitem{GarrigaMukhanov}
J.~Garriga and V.~F.~Mukhanov,
``Perturbations in k-inflation,''
Phys.\ Lett.\ B {\bf 458}, 219 (1999)
[arXiv:hep-th/9904176].

\bibitem{BirrellDavies}  N.~D.~Birrell   and  P.~C.~Davies,  ``Quantum
  Fields  In Curved Space,''  (Cambridge University  Press, Cambridge,
  UK, 1982).
 
\bibitem{polytropic}
S. Chandrasekhar, ``An Introduction to the Study of Stellar Structure,''
(University of Chicago Press, Chicago, USA, 1939). 

\bibitem{k-inflation}
C.~Armendariz-Picon, T.~Damour and V.~Mukhanov,
``k-inflation,''
Phys.\ Lett.\ B {\bf 458}, 209 (1999)
[arXiv:hep-th/9904075].

\bibitem{FinelliBrandenberger}
F.~Finelli and R.~H.~Brandenberger,
``Parametric amplification of metric fluctuations during reheating in 
two  field models,''
Phys.\ Rev.\ D {\bf 62}, 083502 (2000)
[arXiv:hep-ph/0003172].

\bibitem{curvaton} D.~H.~Lyth and D.~Wands, ``Generating the curvature
  perturbation  without an inflaton,''  Phys.\ Lett.\  B {\bf  524}, 5
  (2002) [arXiv:hep-ph/0110002].

\bibitem{ChShTr}
D.~J.~Chung, G.~Shiu and M.~Trodden,
``Running of the scalar spectral index from inflationary models,''
arXiv:astro-ph/0305193.

\end{thebibliography}
\end{document}